\documentclass[preprint,12pt]{aastex}

\shorttitle{Gamma-Ray Burst Trigger Toolkit}

\begin{document}

\title{A Gamma-Ray Burst Trigger Toolkit}
\author{David L. Band\altaffilmark{1}}
\affil{GLAST SSC, Code 661, NASA/Goddard Space Flight Center,
Greenbelt, MD  20771}
\altaffiltext{1}{Joint Center for Astrophysics, Physics
Department, University of Maryland Baltimore County,
1000 Hilltop Circle, Baltimore, MD 21250}
\email{dband@lheapop.gsfc.nasa.gov}

\begin{abstract}
The detection rate of a gamma-ray burst detector can be
increased by using a count rate trigger with many
accumulation times $\Delta t$ and energy bands $\Delta E$.
Because a burst's peak flux varies when averaged over
different $\Delta t$ and $\Delta E$, the nominal
sensitivity (the numerical value of the peak flux) of a
trigger system is less important than how much fainter a
burst could be at the detection threshold as $\Delta t$ and
$\Delta E$ are changed. The relative sensitivity of
different triggers can be quantified by referencing the
detection threshold back to the peak flux for a fiducial
value of $\Delta t$ and $\Delta E$. This mapping between
peak flux values for different sets of $\Delta t$ and
$\Delta E$ varies from burst to burst. Quantitative
estimates of the burst detection rate for a given detector
and trigger system can be based on the observed rate at a
measured peak flux value in this fiducial trigger.
Predictions of a proposed trigger's burst detection rate
depend on the assumed burst population, and these
predictions can be wildly in error for triggers that differ
significantly from previous missions.  I base the fiducial
rate on the BATSE observations: 550 bursts per sky above a
peak flux of 0.3~ph~cm$^{-2}$~s$^{-1}$ averaged over
$\Delta t$=1.024~s and $\Delta E$=50--300~keV. Using a
sample of 100 burst lightcurves I find that triggering on
all possible values of $\Delta t$ that are multiples of
0.064~s decreases the average threshold peak flux on the
1.024~s timescale by a factor of 0.6. Extending $\Delta E$
to lower energies includes the large flux of the X-ray
background, increasing the background count rate.
Consequently a low energy $\Delta E$ is advantageous only
for very soft bursts. Whether a large fraction of the
population of bright bursts is soft is disputed; the new
population of X-ray Flashes is soft but relatively faint.
\end{abstract}


\section{Introduction}

The goal of the next generation of gamma-ray burst missions such
as {\it Swift} and {\it EXIST} is to detect and localize the
largest number of bursts possible. This will be accomplished both
by increasing the detector's effective area or decreasing its
background, and by increasing the sensitivity of the trigger
system.  Increasing the sensitivity through hardware can be
expensive, often requiring a larger and heavier detector.
However, increasing the sensitivity through a more sophisticated
trigger is relatively inexpensive, particularly with the
availability of more powerful flight-qualified CPUs.  But
different triggers respond to different burst properties, and the
relevant question which I address is how much fainter are the
bursts at threshold for the more sophisticated triggers relative
to the previous generation of triggers. I reference different
triggers to a fiducial trigger; normalizing the detection rate for
this fiducial trigger permits estimates of the detection rate for
proposed detector and triggers systems.  However, the actual
detection rate will depend on the burst population, which will be
unknown in unexplored regions of parameter space; consequently an
estimate of a proposed detector's capabilities should state the
assumed burst population.

In a count rate trigger, the detector triggers if the number of
counts accumulated over a given energy band $\Delta E$ and time
scale $\Delta t$ increases by a statistically significant amount
over the expected number of background counts.  The expected
number of background counts is calculated by accumulating counts
over non-burst time intervals.  Current triggers (e.g., for {\it
HETE-II} and {\it Swift}) estimate trends in the background, which
improves the accuracy of the background determination, but does
not increase the sensitivity. The trigger threshold can be
converted into the burst's peak flux (photons s$^{-1}$ cm$^{-1}$)
using the Instrument Response Function (IRF) and a typical burst
spectrum. A more sophisticated trigger may result in a nominally
much greater sensitivity---numerically smaller peak flux at the
trigger threshold---than a simpler trigger.  However, the peak
flux values measured by the different triggers will be over
different energy bands and will be averaged over different
timescales; since bursts have temporal and spectral structure,
they have very different peak fluxes when averaged over these
energy and time bins. Consequently, greater quantitative
sensitivity does not mean that proportionately fainter bursts are
detected. The relevant question is how much fainter are bursts at
threshold for one trigger relative to another trigger.

Here I consider only single-stage count rate trigger systems where
an increase in a single count rate time series suffices. Thus, I
do not consider the sensitivity of triggers where the count rate
trigger must be corroborated by a point source in an imaging
system (as for {\it Swift}); the second stage decreases the
overall sensitivity since the imaging system will reject events
that triggered the count rate trigger system. Nor do I consider
the sensitivity for triggers based on multiple detectors (e.g.,
the WFC and GRBM on {\it BeppoSAX}---Frontera et al. 2000).

The sensitivity of different trigger systems can be
compared by reference to a fiducial system.  Furthermore,
normalizing the rate for the fiducial trigger permits the
estimate of a mission's detection rate, which is required
both to forecast the achievable scientific results and to
plan mission operations (e.g., telemetry or onboard memory
requirements).  I use the results of the Burst And
Transient Source Experiment (BATSE) on the {\it Compton
Gamma-Ray Observatory (CGRO)}, which provided a very large
burst sample accumulated with a well-understood trigger
system

I first present in \S 2 a fiducial trigger system normalized by
the BATSE results.  Next, in \S 3 I formulate a methodology for
evaluating the sensitivity of different triggers relative to the
fiducial trigger, which I then apply to triggers with different
time (\S 4) and energy (\S 5) bins.  A significant issue is the
population of soft bursts (or X-ray flashes) that a detector
sensitive at low energy might detect (\S 6).  The results are
summarized in \S 7.

\section{BATSE Fiducial}

BATSE consisted of 8 modules, each with two different detectors
(Fishman et al. 1989): a large (2025~cm$^2$), flat Large Area
Detector (LAD); and a smaller (127 cm$^2$), thicker Spectroscopy
Detector (SD). The detectors relevant to burst detection, the LADs
were oriented on the faces of a regular octahedron.  BATSE as a
whole triggered when two or more LADs triggered.  The standard LAD
trigger required at least a 5.5$\sigma$ increase in the counts in
the 50--300~keV band accumulated on timescales of 0.064, 0.256 or
1.024~s. Note that this trigger criteria had to be met by the
second most brightly illuminated detector, which was considered by
the BATSE team in calculating their instrument's sensitivity.

The revised 4B catalog (Paciesas et al. 1999) found that BATSE was
0.5 and $\sim0.82$ complete for peak fluxes of 0.25 and
0.3~ph~cm$^{-2}$~s$^{-1}$, respectively, accumulated in 1.024~s
time bins over the 50--300~keV band; I use
0.3~ph~cm$^{-2}$~s$^{-1}$ as BATSE's threshold.  Note that the
conversion from the observed peak count rate to the peak flux
assumes a spectral shape.  The peak flux may have occurred during
a gap in the burst lightcurve, and consequently bursts with such
gaps cannot be used for studies of the peak flux distribution.
Thus the distributions the BATSE team published after they
improved the sky exposure calculation (e.g., Paciesas et al.) used
only bursts without data gaps. However, such bursts occurred, and
must be included in normalizing the burst rate.  In the 4B catalog
there are $\sim 750$ bursts without data gaps above
0.3~ph~cm$^{-2}$~s$^{-1}$. The sky exposure for the 4B~catalog
while BATSE triggered on the 50--300~keV band was 1.73 sky-years.
W.~Paciesas (2002, personal communication) estimates that
$\sim$20\% of the bursts have data gaps. Consequently there are
$\sim 550$ bursts per year per sky above the threshold of
0.3~ph~cm$^{-2}$~s$^{-1}$. This is consistent with the estimate of
$\sim 666$ bursts per year per sky above BATSE's threshold
(Paciesas et al. 1999) after accounting for the bursts BATSE
triggered on with a peak flux below 0.3~ph~cm$^{-2}$~s$^{-1}$
(Paciesas, 2002, personal communication).  At this threshold peak
flux, the BATSE cumulative intensity distribution is approximately
a power law with an index of -0.8.

Was BATSE's threshold indeed at a peak flux of
0.3~ph~cm$^{-2}$~s$^{-1}$? BATSE's sensitivity can be estimated
analytically.  I attribute most of the background below 300~keV to
the X-ray Background (using the formulation of Gruber 1992), and
use the BATSE LAD effective area curve (Fishman et al. 1989) to
estimate the counts from the burst. I find in the 50--300~keV band
a 5.5$\sigma$ threshold of $\sim$0.17-0.18~ph~cm$^{-2}$~s$^{-1}$
on axis. The least sensitive case is when the burst is along the
normal to one detector and thus $\cos \theta=1/3$ for the second
most brightly illuminated detector; the resulting threshold flux
is 0.525~ph~cm$^{-2}$~s$^{-1}$. The most sensitive case is when
the burst falls directly between the normal to 2 detectors, for
which $\cos \theta =0.8165$, giving a peak flux of
0.21~ph~cm$^{-2}$~s$^{-1}$. Based on the symmetry of the
octahedron, there is more solid angle near the most sensitive case
than near the least sensitive case. Additional complications that
are not modelled include the splash off the Earth's atmosphere and
off of the rest of {\it CGRO}, both of which increased the signal
for a given burst flux. Thus the BATSE team's threshold
distribution curve is reasonable.

\section{Trigger Sensitivity}

Assume the number of background counts accumulated on a
time scale $\Delta t$ over an energy range $\Delta E$ is
$B(\Delta t,\Delta E)$.  Here I consider triggers that only
use contiguous accumulation times.  The burst signal---the
increase in the count rate resulting from the burst---is
$S(\Delta t,\Delta E)$. Both $B$ and $S$ are the counts the
detector assigns to the range of energy channels labelled
$\Delta E$, i.e., $B$ and $S$ result from folding the
burst's spectrum through the detector's response.  Thus the
nominal $\Delta E$ will correspond to somewhat different
actual energy ranges for different detectors based on the
detectors' energy response. I do not consider the apparent
increase in $B$ that can occur when burst flux is included
in the estimate of $B$ if the burst rises slowly before the
peak flux. The significance of $S$ is
\begin{equation}
\sigma(\Delta t,\Delta E) = S(\Delta t,\Delta E) /
   \sqrt{B(\Delta t,\Delta E)} \quad .
\end{equation}
Since $B(\Delta t,\Delta E)$ is determined primarily by the
design of the detector (although there may also be a
dependency on time variable particle fluxes, internal
activation, etc.), specifying the threshold value of
$\sigma$, $\sigma_0$, sets the minimum value $S_0(\Delta
t,\Delta E)$ that triggers the detector. This value of
$S_0$ is used as a measure of the detector's sensitivity.

However, $S(\Delta t,\Delta E)$ is a function of $\Delta t$
and $\Delta E$ for a given burst, and thus the threshold
bursts for triggers using different sets of $\Delta t$ and
$\Delta E$ are not fainter in proportion to the threshold
values $S_0$ for the different sets.  Therefore, I compare
the significance for a given set of $\Delta t$ and $\Delta
E$ to the significance for a fiducial set $\Delta t_f$ and
$\Delta E_f$ for a given burst,
\begin{equation}
R(\Delta t,\Delta E;\Delta t_f,\Delta E_f) =
   {{\sigma(\Delta t,\Delta E)}\over{\sigma(\Delta t_f,\Delta E_f)}}
   ={{S(\Delta t,\Delta E)}\over {S(\Delta t_f,\Delta E_f)}}
   \sqrt{{B(\Delta t_f,\Delta E_f)}\over {B(\Delta t,\Delta E)}}
   \quad .
\end{equation}
$R$ is the factor by which a given burst could have been
fainter and still be detected by a trigger with $\Delta t$
and $\Delta E$ relative to a trigger with the fiducial set
$\Delta t_f$ and $\Delta E_f$.  Clearly, a more sensitive
detector and trigger system has a greater value of $R$, and
for a fixed threshold $\sigma_0$, $S_0 \propto 1/R$. Since
$S(\Delta t,\Delta E)$ varies from burst to burst, $R$ will
vary from burst to burst, resulting in a distribution of
$R$ values for the burst ensemble.

Therefore, to estimate the detection rate for a detector with a
prescribed trigger system the properties of the burst ensemble
must be known or assumed.  Conversely, assumptions about the
prevalence of bursts with spectral or temporal properties to which
previous trigger systems were relatively insensitive (e.g., the
rate of bursts with $E_p=10$~keV---$E_p$ is the peak of the
$E^2N_E\propto \nu F_\nu$ spectrum---to which BATSE's 50--300~keV
trigger was insensitive) will have a large effect on the estimated
burst rate for trigger systems with very different sensitivities.
Consequently the assumptions about the burst population must
accompany any estimate of a detector's burst detection rate.

Burst samples are often defined by intensities related to the
trigger because the completeness of the observed intensity
distribution is understood. For example, the BATSE intensity
distribution was usually presented as log~$N$-log~$P$, where $P$
was the peak flux in the usual BATSE $\Delta E=50$--300~keV over
one of the three values of $\Delta t$.  The statistical study of
burst samples accumulated by detectors with complicated trigger
systems with many values of $\Delta E$ and $\Delta t$ should
nonetheless characterize the bursts with a simple definition of
the intensity, such as the peak flux over a single set of $\Delta
E$ and $\Delta t$. For statistical analysis it may be necessary to
consider only those bursts above a threshold value of the
intensity chosen to characterize the bursts, perhaps discarding a
large fraction of the detected bursts.

\section{Sensitivity to Temporal Structure}

Here I am concerned only with the dependence on the accumulation
time $\Delta t$, and therefore I drop the explicit dependence on
$\Delta E$.  Note that the dependencies on $\Delta t$ and $\Delta
E$ are not separable since burst light curves differ by energy
band. Next, I assume that the background does not vary
significantly over the burst, and therefore $B\propto \Delta t$
($B$ and $S$ are the number of background and source counts
accumulated over $\Delta t$). Consequently $\sqrt{B(\Delta t_f)/
B(\Delta t)} = (\Delta t / \Delta t_f)^{-1/2}$.  The burst counts
$S$ are characterized by two extremes.  The burst may be a pulse
much shorter than both $\Delta t$ and $\Delta t_f$ and therefore
$S(\Delta t)/S(\Delta t_f)\simeq 1$. Alternatively, the burst may
be flat-topped with a peak count rate that is constant over a time
greater than both $\Delta t$ and $\Delta t_f$, and consequently
$S(\Delta t)/S(\Delta t_f)=\Delta t / \Delta t_f$. Thus $R(\Delta
t;\Delta t_f)$ is bracketed by $(\Delta t / \Delta t_f)^{\pm
1/2}$.  A trigger with a different $\Delta t$ can be either more
or less sensitive than the fiducial trigger!

One can use a very large number of accumulation times, not just
BATSE's $\Delta t=0.064$, 0.256 and 1.024~s.  Note that BATSE used
consecutive non-overlapping time accumulation bins, and thus the
registration of the burst with respect to these time bins
determined whether a weak burst triggered the detector; the peak
with the maximum count rate may have fallen into either one or two
accumulation bins. The ``ultimate'' trigger would use
accumulations over all possible overlapping accumulation times.
The maximum sensitivity from this ``ultimate'' trigger is the
maximum value of $R$ with respect to $\Delta t$.

To determine the distribution of $R$ for realistic burst light
curves, I use the BATSE 0.064~s resolution light curves for
channels 2+3 ($\sim 50$--300~keV) from all the LADs that
triggered; these light curves are available at
ftp://cossc.gsfc.nasa.gov/compton/data/batse/ ascii\_data/64ms/.
The burst sample consists of the 100 BATSE bursts for which the
data are available with the largest values of $P_{0.064}$, the
peak flux on the 0.064~s time scale. The bursts in this sample are
far above BATSE's detection threshold; they are chosen to
represent fiducial light curves. Of course, systematic differences
between the brightest and dimmest bursts are ignored. I fit a
linear background to these light curves, and then compute $\sigma$
for all possible values of $\Delta t$ that are multiples of
0.064~s. Figure~1 shows the distribution of maximum $R$ using the
most sensitive BATSE trigger; in this case I calculate the peak
value of $R$ with each of BATSE's three values of $\Delta t$
(0.064, 0.256, and 1.024~s) as the fiducial $\Delta t_f$ and use
the minimum resulting value of $R$. Figure~2 shows the
distribution of the peak value of $R$ with the fiducial $\Delta
t_f=1.024$~s. As can be seen, on average the ``ultimate'' trigger
does not increase the significance by large factors over BATSE.
The fiducial ``BATSE'' trigger in these two figures is somewhat
more sensitive than was the actual trigger because I consider all
possible overlapping 0.256 and 1.024~s time bins whereas BATSE
used consecutive, non-overlapping time bins. Figure~3 shows the
distribution of the accumulation times $\Delta t$ that maximize
$R$. This study uses trigger times with a resolution of 0.064~s;
trigger times with a much greater resolution, e.g., 0.01~s, will
greatly increase the sensitivity to very short bursts, and will
increase the sensitivity to all bursts slightly since the
accumulation time can be better tailored to the actual light
curves. However, note that there are few bursts in my study for
which the sensitivity is greatest at the smallest possible
accumulation time of 0.064~s. Thus a much smaller trigger time of
0.01~s may not detect a significantly larger number of bursts.

While choosing an optimum set of $\Delta t$ values is not a goal
of this study, Figure~3 suggests that including triggers with
$\Delta t$ greater than 1~s could increase the burst detection
rate significantly. In particular, $\sim 1/2$ the bursts have
their greatest detection significance for $\Delta t>4$~s. However,
triggers with very long values of $\Delta t$ may be compromised by
trends in the background.  A trigger with $\Delta t<0.064$~s would
be more sensitive to short bursts than BATSE's shortest trigger of
$\Delta t=0.064$~s. Thus triggers spaced logarithmically with
$\Delta t$ between $\sim0.01$~s and $\sim20$~s should be
considered.  Note that Figure~3 uses strong bursts in the
50--300~keV band; the lightcurves will be systematically different
in other energy bands and for fainter bursts in the same energy
band.

\section{Sensitivity to Energy Band}

Next I consider the energy band dependence, and hold the
accumulation time constant.  Consequently, I drop the explicit
dependence on $\Delta t$.  Thus
\begin{equation}
R(\Delta E;\Delta E_f) =
   {{S(\Delta E)}\over {S(\Delta E_f)}}
   \sqrt{{B(\Delta E_f)}\over {B(\Delta E)}}
   \quad .
\end{equation}
The background is usually dominated by the X-ray background below
$\sim 200$~keV. The sensitivity involves a competition between the
burst spectrum and the diffuse background. Since the spectrum
changes from burst to burst, the trigger energy range $\Delta E$
that maximizes $R$ varies.

As an example, I calculate $R$ for all possible energy bands with
edges at 10, 20, 30, 40, 50, 100, 200, 300 and 500~keV, i.e.,
10--20~keV, 10--30~keV, 20--30~keV, etc. The fiducial energy band
is 50--300~keV, which was BATSE's usual trigger band.  For the
background I use the X-ray background and an approximate internal
background that dominates above $\sim250$~keV.  I parameterize the
burst spectrum with the traditional four parameter GRB spectrum
(Band et al. 1993). I assume the detector's efficiency is constant
as a function of energy. $R$ is a function of the three parameters
which determine the shape of the GRB function---two spectral
indices and a break energy. Figure~4 shows the dependence of $R$
on $E_p$ (the energy at which $E^2N_E\propto \nu F_\nu$ peaks) for
four sets of the power law indices.  As can be seen, only for low
values of $E_p$ is the maximum $R$ significantly larger than for
the fiducial energy band.  The sensitivity also increases for high
values of $E_p$ because the detector efficiency is assumed to
remain high at high energy, which is usually not the case (e.g.,
for Cadmium-Zinc-Telluride, the detector material for {\it Swift}
and {\it EXIST}, although the proposed {\it EXIST} detectors will
be thicker, and thus be more efficient at high energy than {\it
Swift}; {\it EXIST}'s large area of active CsI shielding will
increase its high energy response). A more negative value of the
high energy spectral index $\beta$ (i.e., softer high energy
spectrum) results in a greater sensitivity to low values of $E_p$
when $\Delta E$ extends to low energy because there are relatively
few photons in the fiducial band.

\section{Low Values of $E_p$}

The efficacy of lowering the low energy end of $\Delta E$ depends
on the distribution of $E_p$ for the bursts' peak flux. The
spectral evolution studies (e.g., Ford et al. 1995) found that
$E_p$ is greatest when the count rate peaks early in the burst,
the part of the light curve that usually triggers detectors.
Preece et al. (2000) fit 5500 time-resolved spectra from 156
bursts with various spectral models, including the GRB model (Band
et al. 1993).  The $E_p$ distribution in this study peaks between
200 and 250~keV with very few values below 100~keV.  This study
used LAD spectra starting at 25~keV, and thus should have been
able to fit $E_p$ values lower than were observed. The Preece et
al. sample fit spectra from different parts of the burst, not just
the peak, and the spectra from the peak often were accumulated
over long periods (so that the spectra had a sufficient number of
counts). Thus the Preece et al. distribution underestimates the
$E_p$ of the peak flux.  As an aside, the distributions of the low
and high energy spectral indices peak at $\alpha\sim -0.8$ and at
$\beta \sim -2.3$, respectively.  Similarly, Mallozzi et al.
(1995) fit the spectra accumulated over the entire burst for $\sim
400$ bursts; the $E_p$ for these spectra are lower than the $E_p$
of peak flux. Dimmer bursts have lower $E_p$; nonetheless, the
average for the dimmest 1/5 of the bursts is $\langle E_p \rangle
\sim 175$, and only $\sim15$\% of the bursts in this dimmest group
had $E_p<80$~keV.

On the other hand {\it Ginga} observed burst spectra that were
significantly softer than those found by BATSE; 13 out of 22
bursts analyzed by Strohmayer et al. (1998) had $E_p<80$~keV. The
discrepancy between the BATSE and {\it Ginga} results has never
been explained.  Both {\it Ginga} and BATSE triggered on the same
energy range, and thus should have selected the same burst
population.  Since the {\it Ginga} detectors were much smaller
than BATSE's LADs, the bursts {\it Ginga} studied were drawn from
the bright end of BATSE's burst distribution. {\it Ginga} had a
proportional counter sensitive to $\sim 2$~keV, whereas BATSE's
spectra extended down only to $\sim18$~keV; thus {\it Ginga} would
have been able to detect low energy rollovers to which BATSE would
have been insensitive. However, if many bursts are as soft as {\it
Ginga} observed, BATSE would have seen many steep power law
spectra, but did not.  In addition, BATSE consisted of 8 detectors
of two different types while {\it Ginga} had only one low energy
detector.  The burst direction relative to BATSE was known, while
the burst direction was generally unknown for {\it Ginga}.

Preece et al. (1996) calibrated the DISCSP, the lowest energy
discriminator channel for BATSE's SD detectors, providing an
additional energy channel just below the SHERB spectra, the SD
spectra resulting from pulse height analysis.  They performed
joint fits between the DISCSP (with energies in the 5--20~keV
range, depending on the gain setting in effect at the time of the
burst) and the SHERB data. They also compared the observed count
rate in this DISCSP discriminator channel to the prediction from
fits to the SHERB spectrum. While both data types resulted from
the same detector, the two data types used different electronics.
In 12 out of 86 bursts the observed DISCSP count rate was
$>5\sigma$ higher than the prediction; the difference between the
observed and predicted count rate was attributed to a low energy
excess, and not a softer spectrum. The higher energy SHERB spectra
were well-fit by the ``GRB'' function (Band et al. 1993) with
$E_p>100$~keV. It should be noted that this excess resulted from
the comparison of two different data types, which is always
difficult, and from integrations over the entire burst. Observing
this excess in the same data type would confirm its reality.

Frontera et al. (2000) performed spectral fits to the WFC
(2--26~keV) and GRBM (40--700~keV) data for 8 bursts observed by
{\it BeppoSAX}.  Each burst was broken into a number of time
segments.  In 7 out of the 8 bursts the fit to the time segment
with the peak flux in the GRBM band provided only a lower limit to
$E_p$ which was greater than 170~keV; in four cases $E_p>700$~keV.
Only GRB980425 had $E_p=68\pm 40$~keV at the peak.  The bursts in
this sample had to have significant flux in both the WFC and GRBM
detectors, which introduces complicated selection effects.
Nonetheless, this sample does not suggest a large population of
bright soft gamma-ray bursts.

The newly-identified X-ray Flashes (XRFs---Heise et al. 2001) are
a population of soft transients which may be related to the
classical gamma-ray bursts (Kippen et al. 2002). These XRFs are
detected in {\it BeppoSAX}'s WFC but not the GRBM; Kippen et al.
found that 9 of the 10 XRFs which could have been observed by
BATSE had detectable flux in the untriggered BATSE data. As noted
above, dimmer bursts are softer (Mallozzi et al. 1995), but as a
group, the XRFs have smaller $E_p$ values than predicted by the
$E_p$-peak flux correlation.  Of course, the XRFs are X-ray
selected (by the WFC), and a bias towards X-ray richness is
expected. The physical question is whether the XRFs have the same
origin as the classical bursts; the operational question is the
number of bursts or XRFs a sensitive detector with a low energy
trigger band $\Delta E$ would detect.

{\it HETE-2}'s 6--400~keV FREGATE detector demonstrates with a
single detector the trend indicated by previous missions (Barraud
et al. 2002).  Bright bursts have high $E_p$ values and small
X-ray to gamma ray ratios while faint bursts have low $E_p$ and
are X-ray rich; the trend appears to be continuous between
classical bursts and XRFs. While Barraud et al. do not suspect an
inconsistency between their results and {\it Ginga}'s, the bright
soft bursts reported by {\it Ginga} are absent from their data.
FREGATE triggers on energy bands $\Delta E$ that extend as low as
6~keV, and therefore should be more sensitive to bright soft
bursts than was {\it Ginga}.

In conclusion, the BATSE, {\it BeppoSAX} and FREGATE observations
are inconsistent with the fraction of bright soft bursts observed
by {\it Ginga}.  The BATSE SD low energy discriminator suggests
that there is an additional soft component.  The {\it BeppoSAX}
detections of XRFs indicate there is a population of faint soft
transients.  FREGATE shows a continuous trend from bright
classical gamma ray-rich bursts to faint X-ray rich XRFs.
Therefore sensitive low energy detectors may detect large numbers
of XRFs.

\section{Summary}

More sophisticated triggers can increase the sensitivity of a
detector system.  Here I consider count rate triggers, where a
statistically significant increase in the count rate triggers the
detector.  The proposed triggers use a variety of accumulation
times $\Delta t$, energy bands $\Delta E$, and background rate
estimation methods.

Traditionally (e.g., for BATSE) the background was assumed to be
constant and equal to the average over a period of time.  More
sophisticated triggers (e.g., for {\it Swift}) attempt to estimate
trends in the background by considering the background over
various blocks of time. These more complicated background
calculations should increase the accuracy of the background
estimates, but not the nominal sensitivity of the trigger.

Here I analyze triggers that use contiguous accumulation times and
energy bands.  Greater sensitivity might be gained by excluding
the energies of prominent background lines, or by using the peaks
of the highly variable burst lightcurves (e.g., using Bayesian
blocks to isolate high flux periods); however, I do not consider
such triggers. The most sensitive triggers I consider use all
possible values of $\Delta t$ and $\Delta E$.

A detector's threshold peak flux will differ for different sets of
$\Delta t$ and $\Delta E$.  These peak flux values will be the
averages over these $\Delta t$ and $\Delta E$. Because bursts have
highly variable lightcurves, and evolving spectra, a burst's peak
flux for one set of $\Delta t$ and $\Delta E$ will not be equal to
that for another set.  As a concrete example, a trigger's
threshold peak flux is generally proportional to $\Delta
t^{-1/2}$; thus the threshold peak flux for $\Delta t=4$~s will be
1/2 that for $\Delta t=1$~s.  However, if the burst is less than
1~s long, then the $\Delta t=4$~s trigger will average the burst
over 4~s and will consider the burst to have a peak flux 1/4 that
of the peak flux averaged over $\Delta t=1$~s. Thus the $\Delta
t=4$~s trigger will have a nominal sensitivity twice as good as
the $\Delta t=1$~s trigger, but will trigger on short bursts only
half as faint as the $\Delta t=1$~s trigger!

Multiple values of $\Delta t$ and $\Delta E$ complicate the
definition of complete statistical samples.  Of course, only
samples that describe bursts with the same intensity parameter
(e.g., peak flux averaged over a particular set of $\Delta t$ and
$\Delta E$) should be compared. Statistical samples may require a
threshold value of the chosen intensity parameter, resulting in
the exclusion of a large fraction of the detected bursts.  The
purpose of missions with complicated triggering systems is to
detect a larger number of bursts, not to construct a single
complete statistical sample.

To quantify the relative sensitivities of different sets of
$\Delta t$ and $\Delta E$, I reference these sensitivities to the
fiducial $\Delta t$=1.024~s and $\Delta E$=50--300~keV.  Estimates
of the detection rate of a detector and trigger system can be
estimated from these relative sensitivities and the BATSE-observed
burst rate of $\sim550$ bursts per sky-year at a peak flux of
0.3~photons~cm$^{-2}$~s$^{-1}$ in this fiducial trigger band.
Using the lightcurves of the 100 brightest BATSE bursts shows that
using a trigger time which is any multiple of 0.064~s would have
permitted BATSE to detect bursts a factor of 1.3 fainter. While
the burst photon flux increases as the low end of $\Delta E$ is
pushed down to 10~keV, the photon flux of the diffuse background
increases even more rapidly, and thus only bursts with very soft
spectra with a paucity of photons in the higher energy trigger
band will benefit from the lowering the low end of $\Delta E$.
XRFs are a population of faint soft X-ray transients that
sensitive low energy detectors should detect.

Finally, the predicted burst detection rate of a new detector and
trigger system depends on the assumed burst population, and
estimates justifying a proposed detector should specify the burst
population used to model the detection rate.  An actual detection
rate significantly lower than the prediction should not be
regarded as a failure of the mission, but as a discovery about the
burst population.

\acknowledgements

I would like to thank M.~Briggs, E.~Fenimore, N.~Gehrels,
J.~Grindlay, M.~Kippen, J.~Norris, W.~Paciesas, and R.~Preece for
discussions and comments upon this work.

\clearpage

\begin{figure}
\plotone{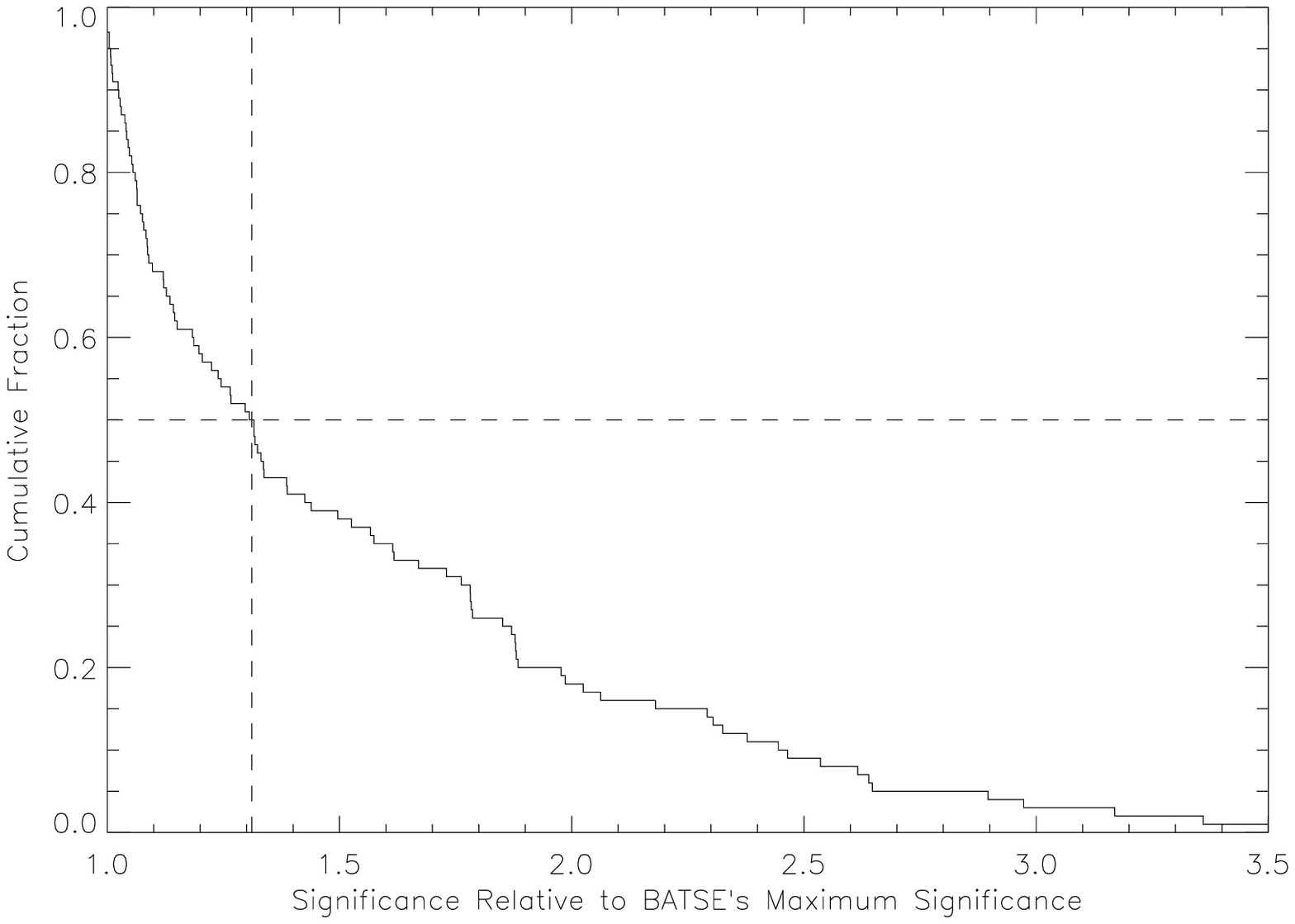} \caption{Cumulative distribution of the maximum
significance relative to the maximum significance on BATSE's three
trigger timescales of $\Delta t=0.064$, 0.0256, and 1.024~s.  A
larger value means that a fainter burst will trigger the detector.
\label{brmin}}
\end{figure}

\begin{figure}
\plotone{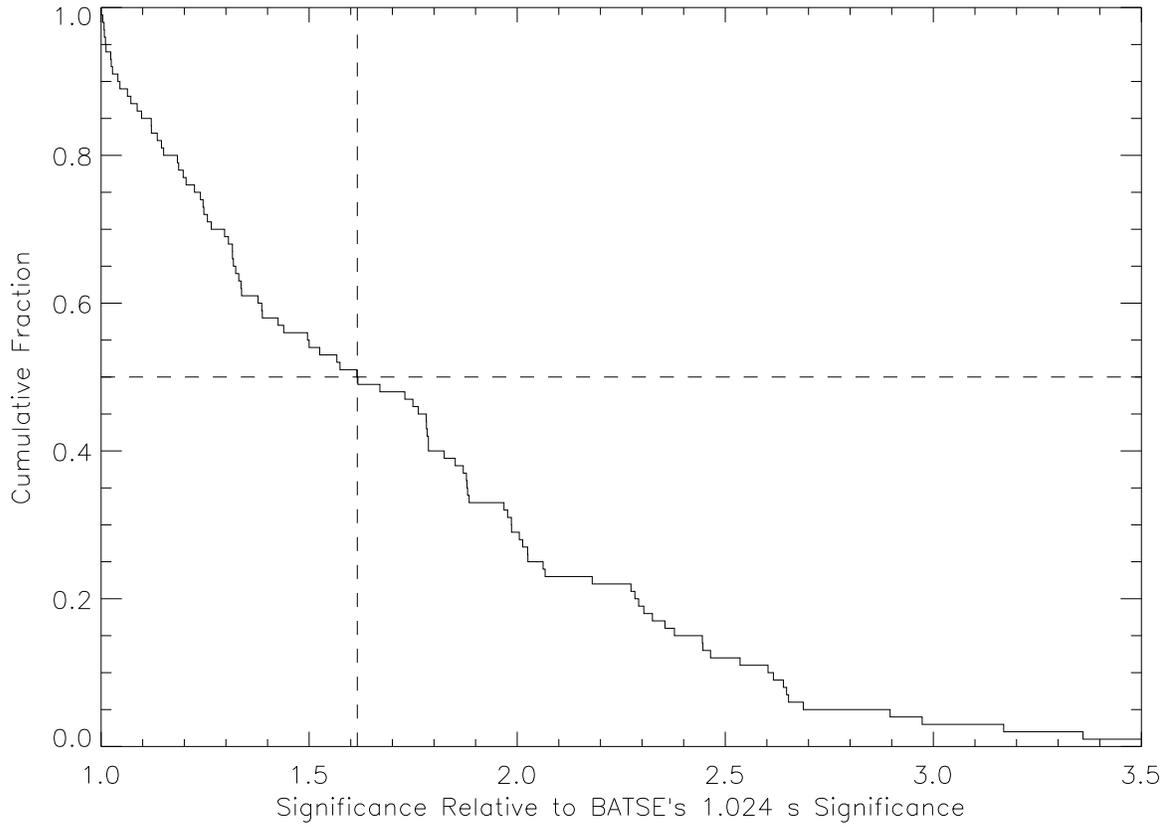} \caption{Cumulative distribution of the
maximum significance relative to the significance on
BATSE's 1.024~s timescale.\label{br1024}}
\end{figure}

\begin{figure}
\plotone{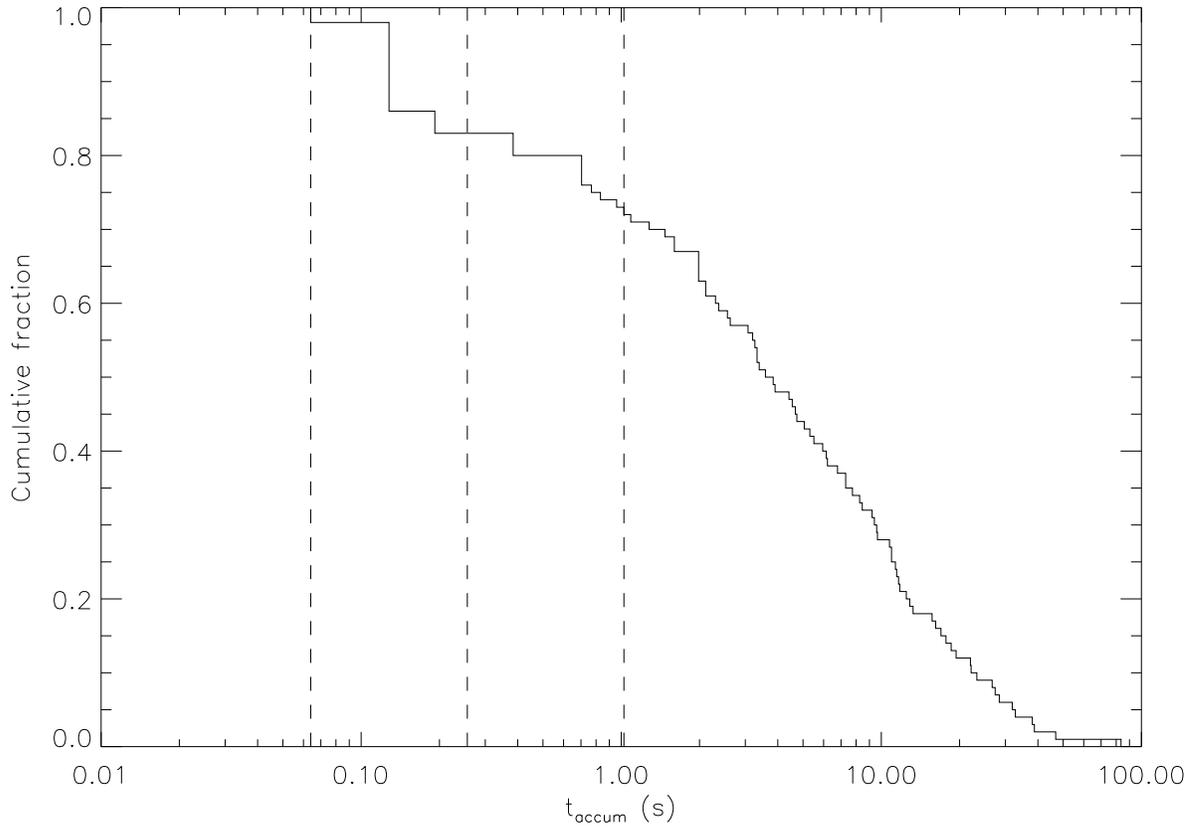}
\caption{Cumulative distribution of the
accumulation time of the maximum significance.\label{tmax}}
\end{figure}

\begin{figure}
\plotone{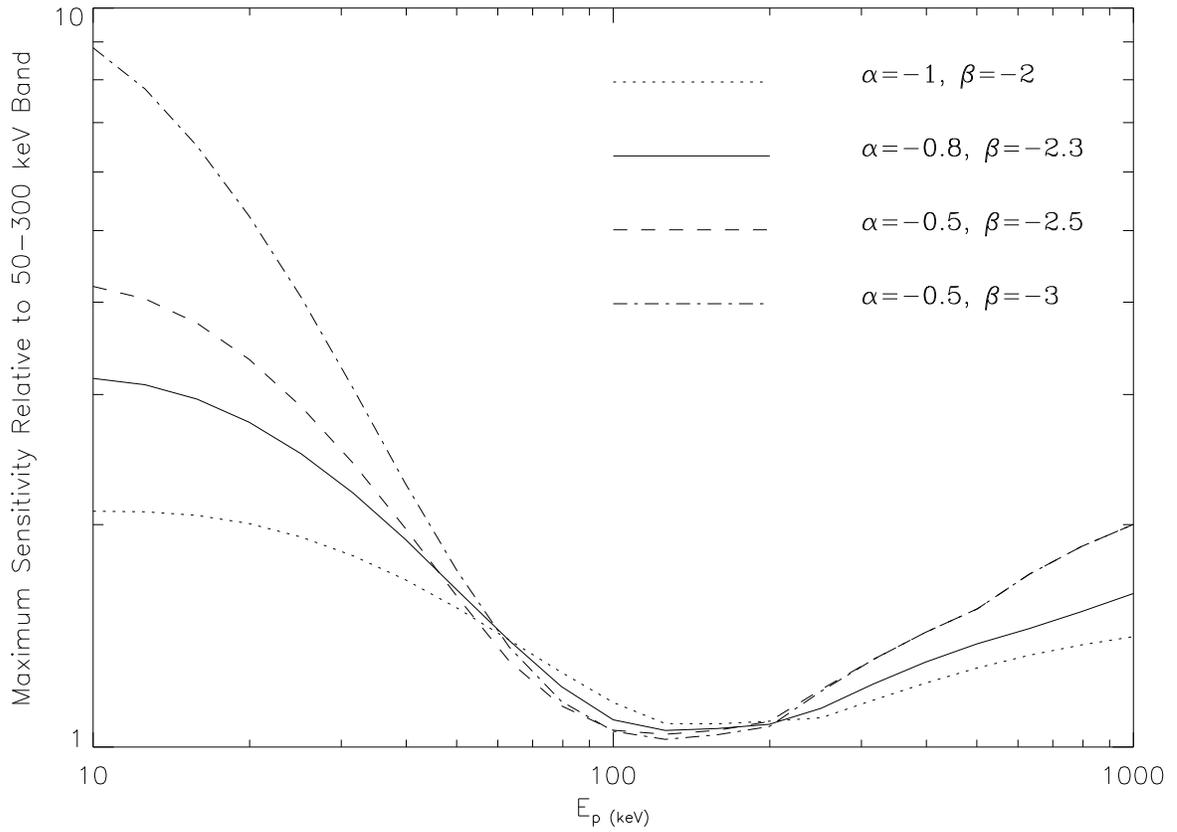} \caption{Sensitivity for the optimum energy band
relative to the 50--300~keV band as a function of $E_p$.  A larger
value means that a fainter burst will trigger the detector.
\label{erange}}
\end{figure}

\end{document}